\documentclass[pdftex,twocolumn,epjc3]{svjour3}          

\RequirePackage[T1]{fontenc}

\smartqed  

\RequirePackage[dvips]{graphicx}
\RequirePackage{mathptmx}      
\RequirePackage{flushend}
\RequirePackage[numbers,sort&compress]{natbib}
\RequirePackage[colorlinks,citecolor=blue,urlcolor=blue,linkcolor=blue]
{hyperref}

\RequirePackage{amssymb}

\journalname{Submitted to Eur. Phys. J. C}

\begin{document}

\title{Procedure for short-lived particle detection in the OPERA experiment \\and its application to charm decays}

\author{
N.~Agafonova\thanksref{MOSCOWINR}\and  
A.~Anokhina\thanksref{MOSCOWSINP}\and  
S.~Aoki\thanksref{KOBE}\and
A.~Ariga\thanksref{BERN}\and
T.~Ariga\thanksref{BERN}\and
D.~Bender\thanksref{ANKARA}\and
A.~Bertolin\thanksref{PADOVAINFN}\and
C.~Bozza\thanksref{SALERNO}\and
R.~Brugnera\thanksref{PADOVA, PADOVAINFN}\and
A.~Buonaura\thanksref{NAPOLI, NAPOLIINFN}
% ...
}

\thankstext[$\star$]{e1}{Corresponding authors. E-mail: marilisa.deserio@ba.infn.it, alessandra.pastore@ba.infn.it}
\thankstext{e3}{Now at Kyungpook National University, Daegu, Korea.}
\thankstext{e4}{Now at Samsung Changwon Hospital, SKKU, Changwon, Korea.}
\institute{
INR - Institute for Nuclear Research of the Russian Academy of Sciences, RUS-117312 Moscow, Russia\label{MOSCOWINR} \and
SINP MSU - Skobeltsyn Institute of Nuclear Physics, Lomonosov Moscow State University, RUS-119991 Moscow, Russia\label{MOSCOWSINP} \and
Kobe University, J-657-8501 Kobe, Japan\label{KOBE} \and
Albert Einstein Center for Fundamental Physics, Laboratory for High Energy Physics (LHEP), University of Bern, CH-3012 Bern, Switzerland\label{BERN} \and
METU - Middle East Technical University, TR-06531 Ankara, Turkey\label{ANKARA} \and
INFN Sezione di Padova, I-35131 Padova, Italy\label{PADOVAINFN} \and
Dipartimento di Fisica dell'Universit\`a di Salerno and ``Gruppo Collegato'' INFN, I-84084 Fisciano (Salerno), Italy\label{SALERNO} \and
Dipartimento di Fisica dell'Universit\`a di Padova, I-35131 Padova, Italy\label{PADOVA} \and
Dipartimento di Fisica dell'Universit\`a Federico II di Napoli, I-80125 Napoli, Italy\label{NAPOLI} \and
INFN Sezione di Napoli, 80125 Napoli, Italy\label{NAPOLIINFN}
% ...
}

\maketitle

\begin{abstract}

The OPERA experiment, designed to perform the first 
observation of $\nu_\mu \rightarrow \nu_\tau$ oscillations 
in appearance mode through the detection of the $\tau$ leptons produced in $\nu_\tau$ charged current interactions, has collected data from 2008 to 2012. 

In the present paper, the procedure developed to detect $\tau $ particle decays, occurring over distances of the order of  $1 \, \rm mm$ from the neutrino interaction point, is described in detail.
The results of its application to the search for charmed hadrons are then presented as a validation of the methods for $\nu_\tau$ appearance detection. 

\end{abstract}

\section{Introduction}
The OPERA experiment \cite{OPERA} searches for $\nu_\mu \rightarrow \nu_\tau$ oscillations in appearance mode in the CNGS beam \cite{CNGS}.  
In a recent paper \cite{NuTau3}, it has reported evidence for $\nu_\tau$ appearance, following the observation of three $\nu_\tau$ candidates in an extended subsample of collected events with respect to the analysis presented in \cite{NuTau1, NuTau2}.

The detection of $\nu_\mu \rightarrow \nu_\tau$ oscillations is based on the distinctive signature of $\nu_\tau$ charged current (CC) interactions with the final-state $\tau$ particle decaying in one (muon, electron or hadron) or three prongs. 

Thanks to the excellent spatial resolution of nuclear emulsions, interleaved with lead plates to form 
a compact modular target, and using the complementary information provided by electronic detectors, event topology and kinematics can be fully reconstructed \cite{OPERA2}. 

Nuclear emulsions are tracking detectors particularly well suited for the study of short-lived particle decays. They have been successfully used in the DONUT experiment \cite{DONUT_det} at FNAL, providing the first experimental evidence of the $\tau$ neutrino \cite{DONUT_res}, and in the CHORUS experiment \cite{CHORUS_det}, which has investigated $\nu_\mu \rightarrow \nu_\tau$ oscillations at large values of $\Delta m^2$ \cite{CHORUS_res} as well as studied $\nu_\mu$-induced charm production, collecting a sample of more than $\rm 2000$ interactions with a charmed particle in the final state \cite{CHORUS_charm}. 

Charmed hadrons have indeed masses and lifetimes similar to those of the $\tau$ lepton and constitute one of the main background sources for oscillation experiments like OPERA. At the same time, charm production represents the most powerful tool to directly test the experiment capability 
of detecting $\tau$ decays, given the alike topology characterising  $\nu_\mu$ CC events with a charmed particle in the final state and oscillated $\nu_\tau$-induced CC interactions. 

The present paper focuses on the analysis procedure 
to detect short-lived particle decays in OPERA. The results of its application to the search for charmed hadrons are then presented as a validation of the procedure for the detection of $\nu_\tau$ appearance. 

After a brief description of the event location (Sect. 2), the methods applied to detect short-lived particles are presented in detail (Sect. 3). The results of the application of the procedure to the search for charmed particle production in the full data sample of CC  events collected in 2008 and 2009 runs as well as in a subsample of the statistics collected in the 2010 run are finally reported and discussed (Sect. 4).

\section{Event location}

The location of neutrino interactions in the OPERA target has been described in details in \cite{OPERA2}. For the sake of clarity, the main features will be briefly summarised here. 

The OPERA target, with a total mass of about $1.2 \, \rm{kt}$, is divided in two sections and has a modular structure consisting of about $\rm 150000$ basic units called \emph{bricks}. 
A brick is made of 57 thin nuclear emulsion films interleaved with 56 $1 \, \rm{mm}$-thick lead plates with transverse sizes of $12.8 \times 10.2 \, \rm{cm^2}$ and dimension along the beam direction of $7.9 \, \rm{cm}$, corresponding to about $\rm 10$ radiation lengths. 
The OPERA emulsion films are made of two $\rm 45 \, \mu m$-thick sensitive layers deposited on each side of a $\rm 205 \, \mu m$ plastic base.
Bricks are arranged in planar structures, called \emph{walls}, with transverse dimensions of about $6.7 \times 6.7 \, \rm{m^2}$. Each wall, hosting $\rm 64 \times 52$ bricks, is coupled to a pair of target tracker (TT) planes, made of plastic scintillator strips of $2.6 \, \rm{cm}$ width.

A muon spectrometer is placed downstream of each of the two target sections. It consists of a dipolar magnet of about $8.75 \times 8 \, \rm{m^2}$, made of two magnetized iron walls and instrumented with planes of bakelite RPC's, inserted between the iron slabs of each wall. Planes of drift tubes with a spatial resolution of $\sim 300 \, \rm{\mu m}$ are placed in front, behind and in between the magnet walls to improve the muon charge and momentum determination.  

A doublet of low background  emulsion films, called \emph{Changeable Sheets} (CS,~\cite{CS}), attached on the downstream face of each brick, acts as an interface between the electronic detectors, with a spatial accuracy of about $\rm 1 \, cm$, and the emulsion/lead target with sub-micrometric resolution.

Once the brick with the highest probability of containing the neutrino interaction point is identified by real-time  electronic detectors data analysis, the corresponding CS are measured by means of high-speed automated optical microscopes ~\cite{ESS1, ESS2, TS}.

In the analysis of the events collected in 2008 and 2009, a brick was developed whenever a track matching some of the  event hits in the electronic detectors was found in the CS.  Starting from the 2010 run, an improved CS trigger has been defined. Currently, a brick is analysed if:
\begin{itemize}
\item{a track compatible with the reconstructed muon within an angular tolerance of $0.06 \, \rm{rad}$ in each projection or two or more tracks converging in the brick are detected, in the case of CC-like events;}
\item{a track matching aligned isolated hits in the electronic detectors or a converging pattern is found, in the case of neutral current (NC)-like events.}
\end{itemize}

The predictions provided by the CS are located in the most downstream part of the brick and followed upstream (\emph{scan-back}) film by film up to the interaction or decay point(s).  
Tracking inefficiencies are recovered by requiring that the search for each track be iterated in $\rm 5$  consecutive emulsion films whenever a track \emph{disappears} in the scan-back process. 

The most upstream track disappearance point is tagged as candidate neutrino vertex point and a \emph{volume-scan} procedure is applied around it to confirm and fully reconstruct the neutrino interaction. A surface of $1 \, \rm cm^2$ is measured in $\rm 5$ films upstream and $\rm 10$ films downstream of the assumed stopping point in order to track all particles within an angular acceptance of $0.6 \, \rm rad$ with respect to the perpendicular to the films. 
Sub-micrometric accuracy is achieved in the reconstruction of  tracks thanks to the exposure of each brick to an optimal flux of high energy cosmic rays before disassembly \cite{cosmics}, allowing relative misalignments and deformations between consecutive emulsion films to be estimated. 

For runs 2008 and 2009, if the measurement of the most probable brick did not allow locating the neutrino interaction, the analysis was immediately extended to less probable bricks. About 1.6 bricks / event were analysed on average. From runs 2010 onwards, a provisional strategy has been adopted in which all most probable bricks are being analysed first. 

The methods described in the next section are then applied to validate the event topology and search for possible decays. 

\section{Decay search procedure}

Two distinct decay topologies can be identified, depending on the relative positions of the interaction and decay vertices. 

If the decay occurs in the same lead plate as or in the first $\rm 45 \, \mu m$-thick emulsion layer downstream of the neutrino interaction, it is defined as \emph{short decay} and daughter tracks can be detected as particles showing a large  impact parameter with respect to the neutrino vertex. 

Otherwise, it is defined as \emph{long decay} and daughter tracks typically appear as \emph{extra-tracks}, not attached to the reconstructed neutrino interaction point and starting downstream of the vertex film\footnote{In the following, the emulsion film immediately downstream of the neutrino interaction point will be referred to as \emph{vertex film}.}.  

\begin{figure*}[t!]
\centering
\begin{tabular}{cc}
    	\includegraphics[scale=0.32]{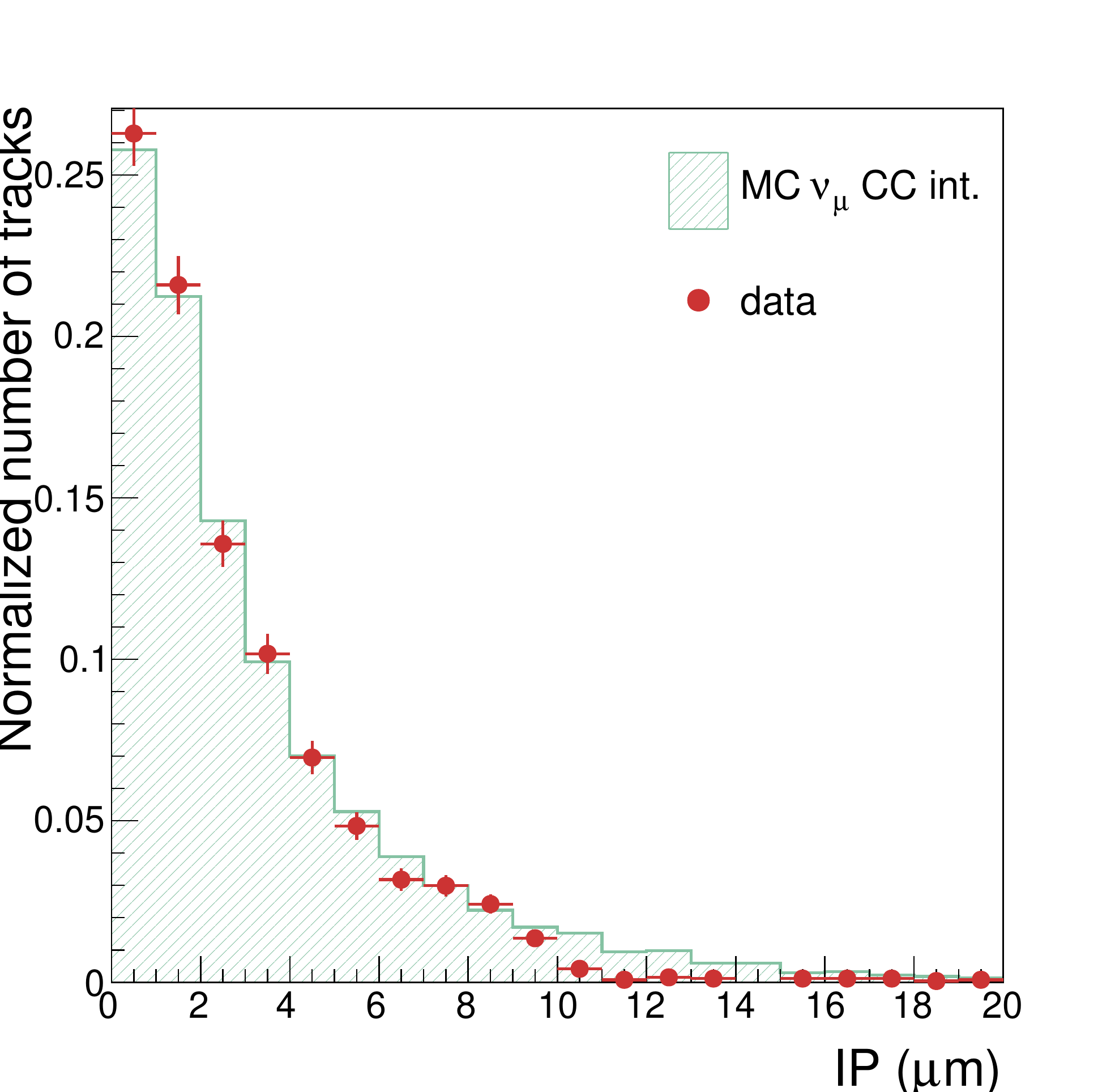}
 & 	\includegraphics[scale=0.32]{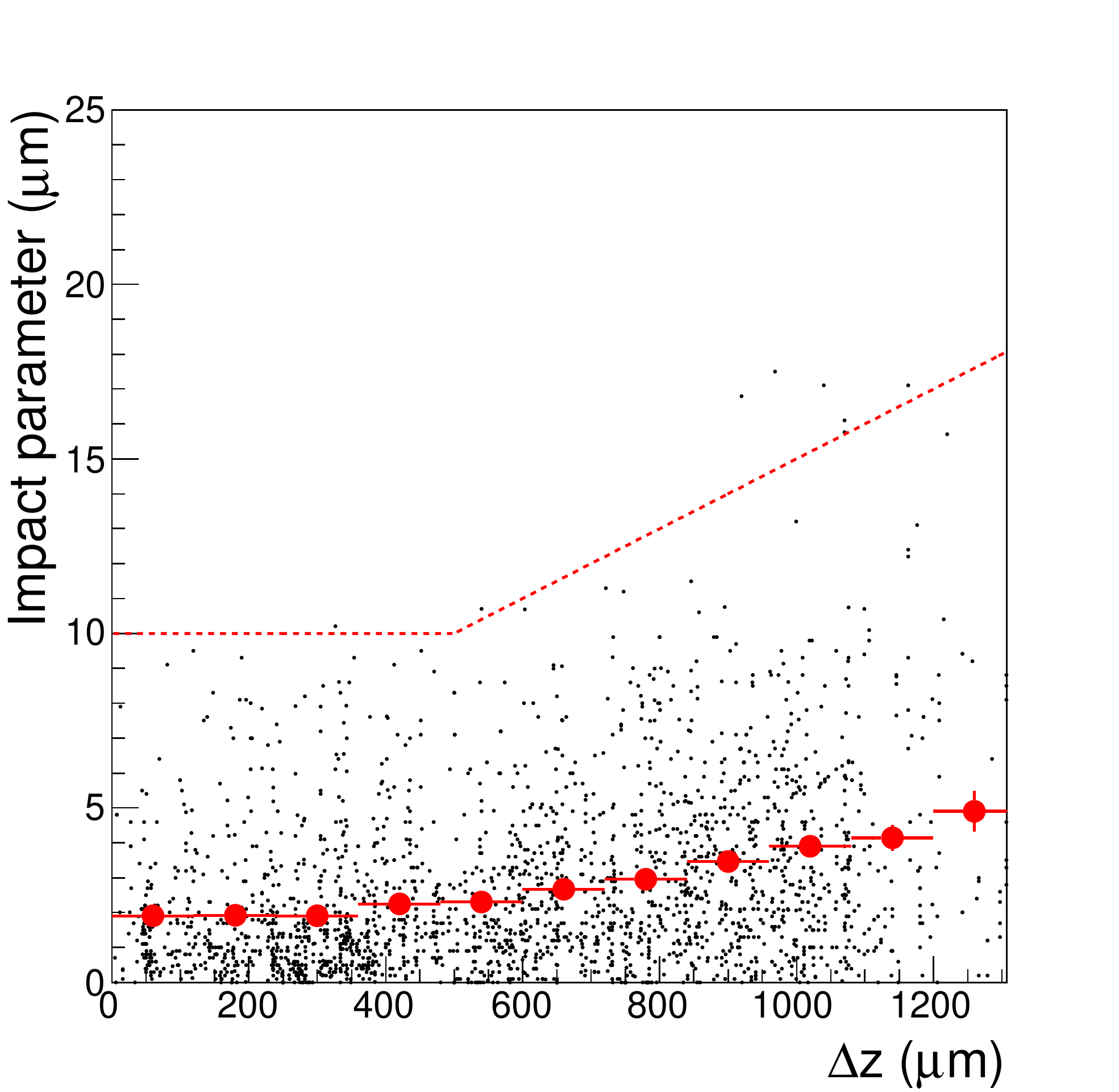}
\end{tabular}
\caption{Left: Impact parameter distribution after the vertex film analysis for a sample of located $\nu_\mu$ CC interactions, compared to Monte Carlo expectations. Right: Track impact parameter as a function of the longitudinal distance from the neutrino interaction vertex (2008-2009 data). The red bullets show the average value for each bin. The dotted red line represents the cut applied to select possible short-decay daughter tracks. \label{fig:IP}}
\end{figure*}

\subsection{Extra-track search}\label{ExtraTrk}

As a result of the standard location procedure, a multi-prong vertex or an isolated track is reconstructed. 
In the latter case, the vertex point is defined as the extrapolation of the track at the center of the lead plate immediately upstream of its disappearance point. 

In order to define the event topology, the first step is to search for additional event-related tracks (extra-tracks), among those reconstructed in the measured volume. 

Only tracks crossing at least three emulsion films and showing an impact parameter with respect to the vertex smaller than $300 \, \rm \mu m$ ($500 \, \rm \mu m)$ are selected, depending on whether the longitudinal distance $\Delta z$ of the most upstream segment of the track with respect to the vertex  is smaller (larger) than $1 \, \rm mm$. Moreover, a cut on the maximum distance from the vertex ($\Delta z \leq 3.6 \, \rm mm$) is applied. 
Indeed, about $99 \, \%$ of the $\tau$ leptons produced in CC interactions of oscillated $\nu_\tau$'s in the OPERA target have a flight lenght smaller than $3.6 \, \rm mm$.

The search for extra-tracks is extended to two emulsion films upstream of the vertex if the event is NC-like: for an event without muon, the possibility that the reconstructed vertex could have been produced by the decay of the $\tau$ lepton created in the final state of a $\nu_\tau$ CC  interaction has to be carefully investigated. A maximum impact parameter of  $500 \, \rm \mu m$ between the extra-track and the vertex is considered. 

Selected extra-tracks are then visually inspected to assess their nature and categorised in:  

\begin{itemize}
\item{$e^+e^-$ pairs from $\gamma$ conversion;}
\item{particles produced in the neutrino interaction and not directly attached to the vertex due to tracking or reconstruction inefficiencies;}
\item{low momentum particles reconstructed as short tracks due to multiple Coulomb scattering;}
\item{secondary particles produced in hadron interactions or decay daughter tracks.}
\end{itemize}

Genuine extra-tracks starting downstream of the vertex film are further analysed and a dedicated \emph{parent search} procedure is applied to detect possible long decays.
The method consists in searching for a track connecting the selected extra-track to the reconstructed vertex with an impact parameter with respect to it smaller than $10 \, \rm \mu m$ and a minimum distance from the possible daughter track smaller than $20 \, \rm \mu m$. Any candidate parent track selected according to these criteria is validated by visual inspection. 

The presence of heavily ionising particles deserves a careful check in order to  discriminate between hadronic interactions and particle decays. 
Heavily ionised tracks are searched for by taking tomographic images around the most upstream segment of the extra-track(s)  or the secondary vertex point.
Images are  automatically processed to identify converging nuclear fragments, typically emitted at large angles 
from the interaction point. 
If no fragment is detected, the vertex is eligible to be a decay.

Extra-tracks starting in the vertex film are analysed as explained in the following section. 

\subsection{Vertex definition and analysis}\label{vtx_analysis}

Tracks directly attached to the reconstructed neutrino interaction point  as well as extra-tracks starting in the vertex film are examined in order to define the event topology and detect possible short decays.

Track segments in the vertex film could be missing in 
the reconstruction because of tracking inefficiency, thus decreasing the resolution in the determination of the neutrino interaction position. Moreover, electron-positron pairs coming from the conversion of $\gamma$'s produced in $\pi^0$ decays and pointing to the vertex can further spoil the accuracy.
By visually inspecting the vertex film, tracking inefficiencies can be recovered and the parameters of all  tracks can be precisely measured and used to re-compute the vertex position. Furthermore, $e^+e^-$ pairs are identified as couples of converging segments and thus tagged and ignored in the determination of the vertex. 

The plot in Figure \ref{fig:IP} left shows the impact parameter distribution after the recovery of missing segments 
in the vertex film and the identification of $e^+e^-$ pairs for a sample of located $\nu_\mu$ CC interactions from 2008, 2009 and 2010 runs, compared to expected Monte Carlo behaviour\footnote{$\nu_\mu$ interactions with charmed particles in the final state are not included in the plot.}. 
The distribution has an average value of about $3 \, \rm \mu m$. Tails are mainly due to low energy particles. 

Figure \ref{fig:IP} right shows the dependence of the track impact parameter on the longitudinal distance between the most upstream measured track segment and the neutrino interaction vertex. 
The dotted red line indicates the applied cut: if a track is found to have an impact parameter larger than $\rm 10 \, \mu m$ for $\rm \Delta z \, \le 500 \, \mu m$ or larger than $\rm (5 \, + \, 0.01 \, \times \, \Delta z) \, \mu m$ for $\rm \Delta z  \, > 500 \, \mu m$, it is further studied in order to assess whether the observed anomalous value can be explained in terms of scattering in the traversed lead thickness. An estimation of the track  momentum is thus performed by comparing the particle slope evolution along the brick with the expected behaviour based on the multiple Coulomb scattering (MCS) \cite{MCS}. Tracks with a measured momentum below $1 \, \rm GeV/c$ are tagged as left by \emph{low momentum} particles and disconnected from the vertex for a more accurate determination of its position. 

If the track length is not sufficient to apply the MCS algorithm, the angular residuals of the track segments in the transverse and longitudinal planes are computed. If both values are above $0.015 \, \rm rad$, the track is classified  as a low momentum particle.

Once the primary vertex has been defined, any track with an anomalous impact parameter that cannot be explained in terms of scattering is carefully investigated and the search for converging tracks and/or nuclear fragments is applied, as explained in Section \ref{ExtraTrk}.

\subsection{Decay search along tracks} \label{InTrack}

Particles having a 1-prong decay (or interaction) topology can be reconstructed as single tracks or as two distinct tracks depending on the kink angle and on the angular resolution which is slope-dependent. In order to enhance the sensitivity to short-lived particles, a search for possible kinks along the tracks attached to the neutrino interaction, not detected in the first stage of the reconstruction, is thus performed in a fiducial volume of 4 films downstream of the vertex.

For each track, the largest angular difference in the 4 most upstream films is computed and compared to the RMS of the  angular deviations along the track. 
If the ratio is larger than 5 and the kink angle, confirmed by direct check in emulsion, is larger than $0.015 \, \rm rad$, the track is classified as potentially interesting and the kink is analysed in order to assess its nature.

\subsection{Validation of decay topologies}

As a result of the procedure illustrated above, events showing a topology compatible with the production and 
decay of a short-lived particle are identified. 
In order to confirm them as candidate $\nu_\tau$ signal events, $\nu_\mu$ interactions with charmed 
particle production and decay or neutrino interactions with hadron re-interaction, a full kinematical analysis is required. 
Details on measurements and discriminating variables used in the analysis can be found in \cite{NuTau2}. 

\section{Charm production in OPERA}

\subsection{Data sample}

In the following, the results of the application of the decay search procedure to the data sample of CC events collected in 2008, 2009 and 2010 runs are presented.

For the first two runs, an inclusive analysis of all predicted  events has been carried out. For the 2010 run, only a subsample of events have been considered including all NC-like events and CC-like events with a reconstructed negative muon with momentum below $\rm 15 \, GeV/c$.\footnote{For the $\nu_\tau$ search, this selection cut has been also applied to the 2008 and 2009 runs \cite{NuTau2, NuTau3} aiming at increasing the signal/noise ratio, with a negligible signal loss.}

The kinematical selection together with the present brick selection strategy described in Section 2 comparatively reduces the charm yield for the 2010 run by about $28 \%$.

The overall data sample consists of $\rm 2925$ $\nu_\mu$ CC located events with the interaction point in the brick\footnote{A fiducial volume cut is applied to remove events with less than 3 emulsion films available for the decay search analysis at the downstream edge of the brick.} and completed decay search.
\subsection{Monte Carlo simulation}

The expected charm yield is computed using the standard  OPERA simulation framework \cite{NuTau2}.  
The neutrino fluxes and spectra are based on a FLUKA \cite{FLUKA} simulation of the CNGS beam-line (2005 revision \cite{Sim_CNGS}). The neutrino interactions in the detector are generated using the  NEGN generator \cite{NEGN}, derived from the NOMAD experiment.

The energy dependence of the charm production cross-section has been obtained starting from the latest results of the CHORUS data analysis \cite{CHORUS_charm} and performing a convolution with the CNGS spectrum. The total charm production 
rate relative to the CC interaction cross-section is $ \sigma (\nu_\mu \, N \, \rightarrow \, \mu \, C \, X) \, / \, \sigma (\nu_\mu \, N \, \rightarrow \, \mu \, X) \, = \, (4.49 \, \pm \, 0.26) \, \%$. The following fragmentation fractions are used in the calculation: $f_{D^+} \, = \, (21.7 \, \pm \, 3.4) \, \%$, $f_{\Lambda_c} \, = \, (25.3 \, \pm \, 4.9) \, \%$, $f_{D_s^+} \, = \, (9.2 \, \pm \, 3.8) \, \%$, $f_{D^0} \, = \, (43.8 \, \pm \, 3.0) \, \%$. 

\begin{table*}\centering
\caption{Decay search efficiencies for the different charmed hadrons. Statistical and systematic errors are summed in
quadrature.}\label{tab:charm_eff}
\begin{tabular}{cccc}
\hline
& \multicolumn{3}{c}{Decay search efficiency} \\
\cline{2-4}
{Particle} & {Short topology} & {Long topology} & {Combined}\\
\hline
{$\rm D^0$} & {$\rm 0.22 \, \pm 0.02$} & {$\rm 0.65 \, \pm 0.05$} & {$\rm 0.39 \, \pm 0.04$}\\
\hline
{$\rm D^+$} & {$\rm 0.20 \, \pm 0.05$} & {$\rm 0.41 \, \pm 0.09$} & {$\rm 0.28 \, \pm 0.05$}\\
\hline
{$\rm D_s^+$} & {$\rm 0.18 \, \pm 0.09$} & {$\rm 0.56 \, \pm 0.30$} & {$\rm 0.33 \, \pm 0.13$}\\
\hline
{$\rm \Lambda_c^+$} & {$\rm 0.10 \, \pm 0.02$} & {$\rm 0.62 \, \pm 0.16$} & {$\rm 0.31 \, \pm 0.07$}\\
\hline
\end{tabular}
\end{table*}

\begin{table*}\centering
\caption{Summary of expected charm and background events compared to observed events. Statistical and systematic errors are summed in quadrature.}\label{tab:charm_res}
\begin{tabular}{lrrrr}
\hline
& \multicolumn{4}{c}{Events} \\
\cline{2-5}
{Decay topology} & \multicolumn{1}{c}{Expected} & \multicolumn{1}{c}{Expected} & \multicolumn{1}{c}{Expected}& {Observed} \\
{} & \multicolumn{1}{c}{charm} & \multicolumn{1}{c}{background} & \multicolumn{1}{c}{total} & {} \\
\hline
{1-prong} & {$\rm 21 \, \pm 2$} & {$\rm 9 \, \pm 3$} & {$\rm 30 \, \pm 4$} & {$19$} \\
\hline
{2-prong} & {$\rm 14 \, \pm 1$} & {$\rm 4 \, \pm 1$} & {$\rm 18 \, \pm 2$} & {$22$} \\
\hline
{3-prong} & {$\rm 4 \, \pm 1$} & {$\rm 1.0 \, \pm 0.3$} & {$\rm 5  \, \pm 1$} & {$5$} \\
\hline
{4-prong} & {$\rm 0.9 \, \pm 0.2$} & {-} & {$\rm 0.9 \, \pm 0.2$} & {$4$} \\
\hline
{Total} & {$\rm 40 \, \pm 3$} & {$\rm 14 \, \pm 3$} & {$\rm 54 \, \pm 4$} & {$50$} \\
\hline
\end{tabular}
\end{table*}

MC-generated events with charmed hadrons in the final state are processed through the full OPERA simulation chain, from the event classification and brick finding provided by the electronics detectors to the CS analysis and event location in the brick, up to the decay search.  

A charmed hadron decay is regarded as \emph{detected} whenever at least one genuine daughter track appears as not attached to the neutrino interaction vertex after the application of the decay search procedure.

Two sources of background are considered in the analysis: hadronic re-interactions and decays of strange particles.

The hadronic background is estimated by using the hadronic particles produced in reconstructed CC events as input to a FLUKA-based MC simulation. The interactions with at least one visible product in emulsion are selected. For 1-prong interactions, only events with the produced hadron showing a momentum greater than $1 \, \rm GeV/c$ and an angle with respect to the interacting hadron larger than $20 \, \rm mrad$ are considered, coherently with the selection criteria applied to experimental data. These cuts reduce the hadronic background by about $\rm 55 \, \%$.

\subsection{Data - MC comparison}

The main contribution to the decay search efficiency comes from the extra-track selection, with an additional improvement of the order of a few percent only coming from the procedure described in Section \ref{InTrack} for long decays. 

The breakdown of the charm detection efficiencies obtained for the different charmed particles is reported in Table \ref{tab:charm_eff}.

The global decay search efficiencies are estimated to be $\rm \epsilon_{short} \, = \, 0.18 \, \pm 0.02 \, (stat) \, \pm 0.01 \, (syst)$ and $\rm \epsilon_{long} \, = \, 0.58 \, \pm 0.07 \, (stat) \, \pm 0.03 \, (syst)$ for short and long topologies respectively. By taking into account the fractions of short and long decays for the charmed species, an overall efficiency $\rm \epsilon_{total} \, = \, 0.34 \, \pm 0.04 \, (stat) \, \pm 0.01 \, (syst)$ is obtained. \newline
\indent A detailed comparison between experimental data and MC expectations is reported in Table \ref{tab:charm_res}. 
The number of observed events showing a topology compatible with that of a $\nu_\mu$ CC interaction with production and decay of a charmed hadron is compared to the expected charm yield and the estimated background for different values of the reconstructed charged particle multiplicity at the secondary vertex. \newline
\indent 
The systematic uncertainties taken into account in the computation, summed in quadrature with the statistical errors in the table, arise from the knowledge of the charm production cross-section, the fragmentation fractions and the branching ratios of the charmed particles in the different decay channels. \newline
\indent The table includes migration among decay modes. As an example, for 1-prong short decays only about $\rm 30 \, \%$ of the expected charm events show a genuine topology, about $\rm 68 \, \%$ of the migrated events being due to the $\rm D^0$  meson. 
The main factors determining a decrease in the observed decay multiplicity with respect to the true multiplicity are found to be the decay search procedure cuts, the tracking inefficiency that can prevent the reconstruction and/or the selection of daughter tracks and the angular acceptance of the volume scan\footnote{For $\nu_\tau$ candidates, dedicated measurements are applied to avoid migration effects, e.g. the angular acceptance in the volume-scan is extended from $\rm 0.6 \, rad$ to $\rm 1 \, rad$.}. In particular, for short topologies, daughter tracks showing impact parameters with respect to the neutrino interaction vertex smaller than $\rm 10 \, \mu m$ can be wrongly attached to it. 
An increase in the reconstructed decay  multiplicity can be  also observed  in a small fraction of events where particles directly attached to the neutrino interaction vertex can be wrongly associated to short decay daughters. \newline 
\begin{figure}
\begin{minipage}{\columnwidth}
\centering
    	\includegraphics[scale=0.27]{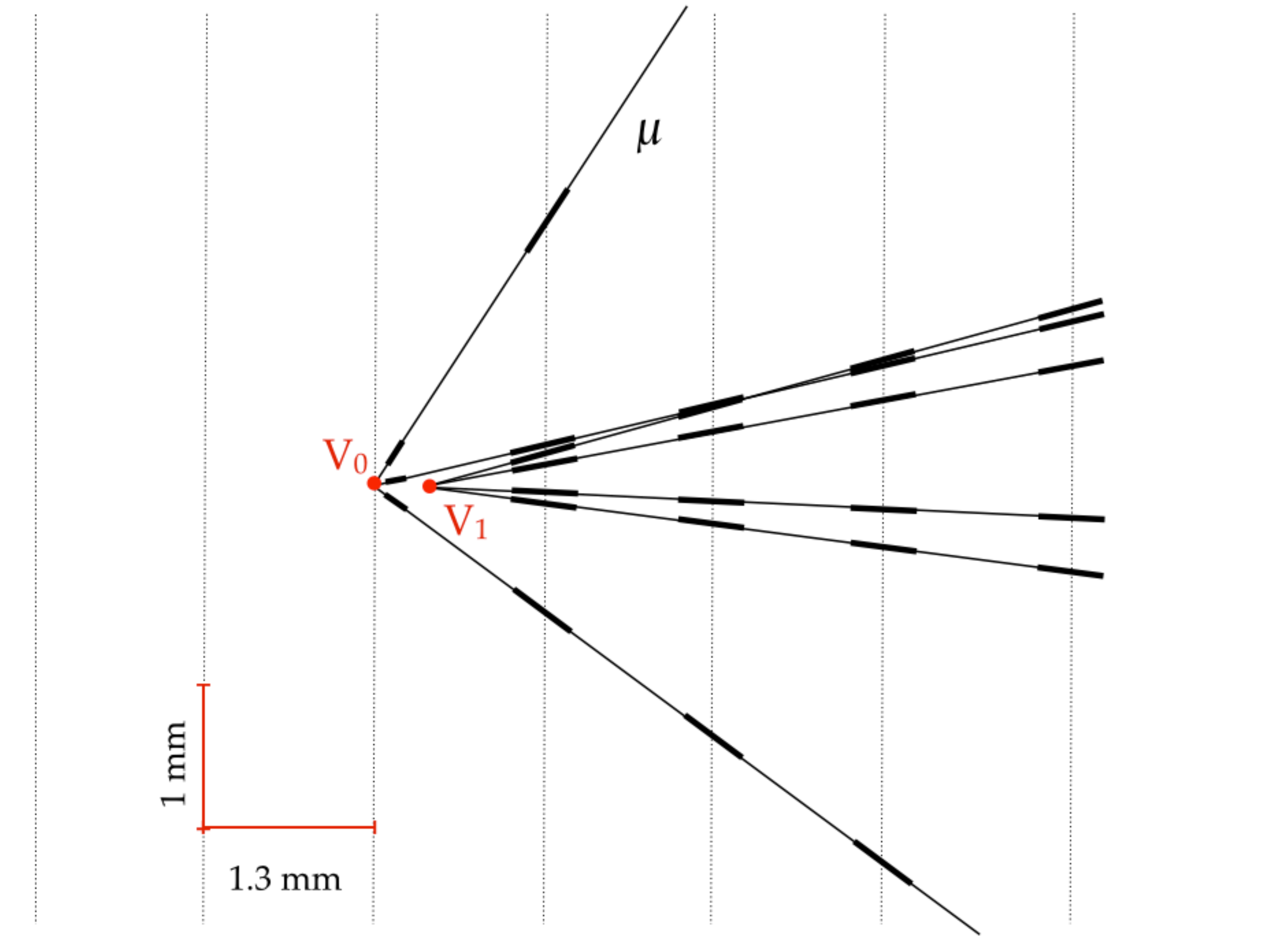}
\caption{Sketch of a reconstructed $\nu_\mu$ CC interaction with a candidate charmed hadron observed in the final state.\label{fig:charm_D0}}
\end{minipage}
\end{figure}
\indent In total, ($\rm 40 \, \pm 3$) charm and ($\rm 14 \, \pm 3$)  background events are expected in the analysed sample, while 
$50$ decay candidate events are observed.  

Hadronic re-interactions constitute the main source of background (about $87 \, \%$ of the total). The decays of strange particles mostly contribute to the 2-prong topology. 

Figure \ref{fig:charm_D0} shows an example of a reconstructed $\nu_\mu$ CC interaction recorded in the run 2008 with a candidate $\rm D^0$ particle in the final state decaying in $\rm 4$ prongs after a distance of about $\rm 310 \, \mu m$. 
The muon track and the candidate charmed meson form an angle of about $\rm 170^\circ$ in the plane transverse to the neutrino direction, thus lying in a back-to-back configuration as expected for charm production. The minimum invariant mass \cite{min_inv_mass} is estimated to be $\rm 1.7 \pm 0.4 \, GeV/c^2$ and is consistent with that of the $\rm D^0$ particle.

\begin{figure*}[h!]
\centering
\begin{tabular}{cc}
	\includegraphics[scale=0.32]{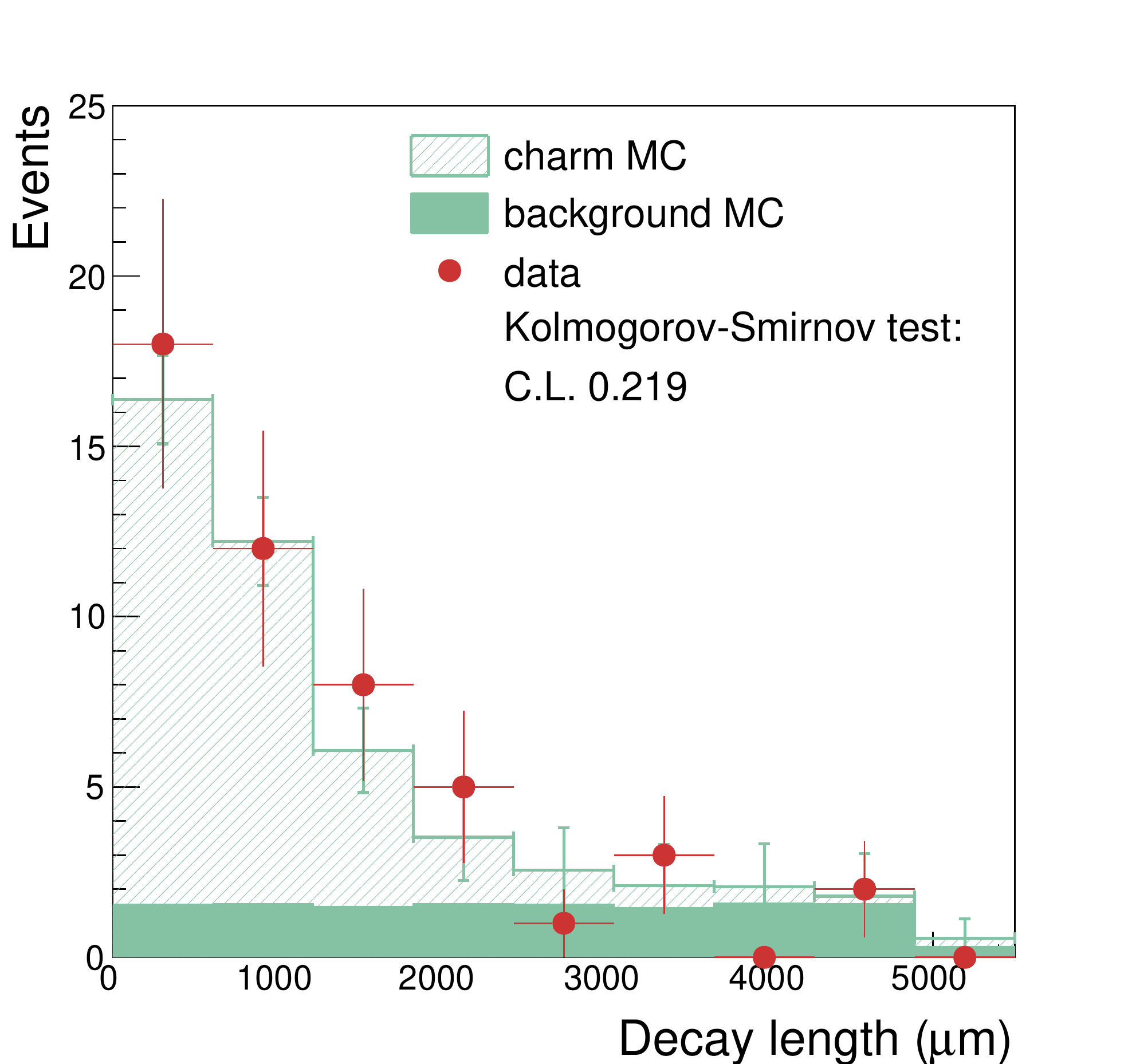}
 & 	\includegraphics[scale=0.32]{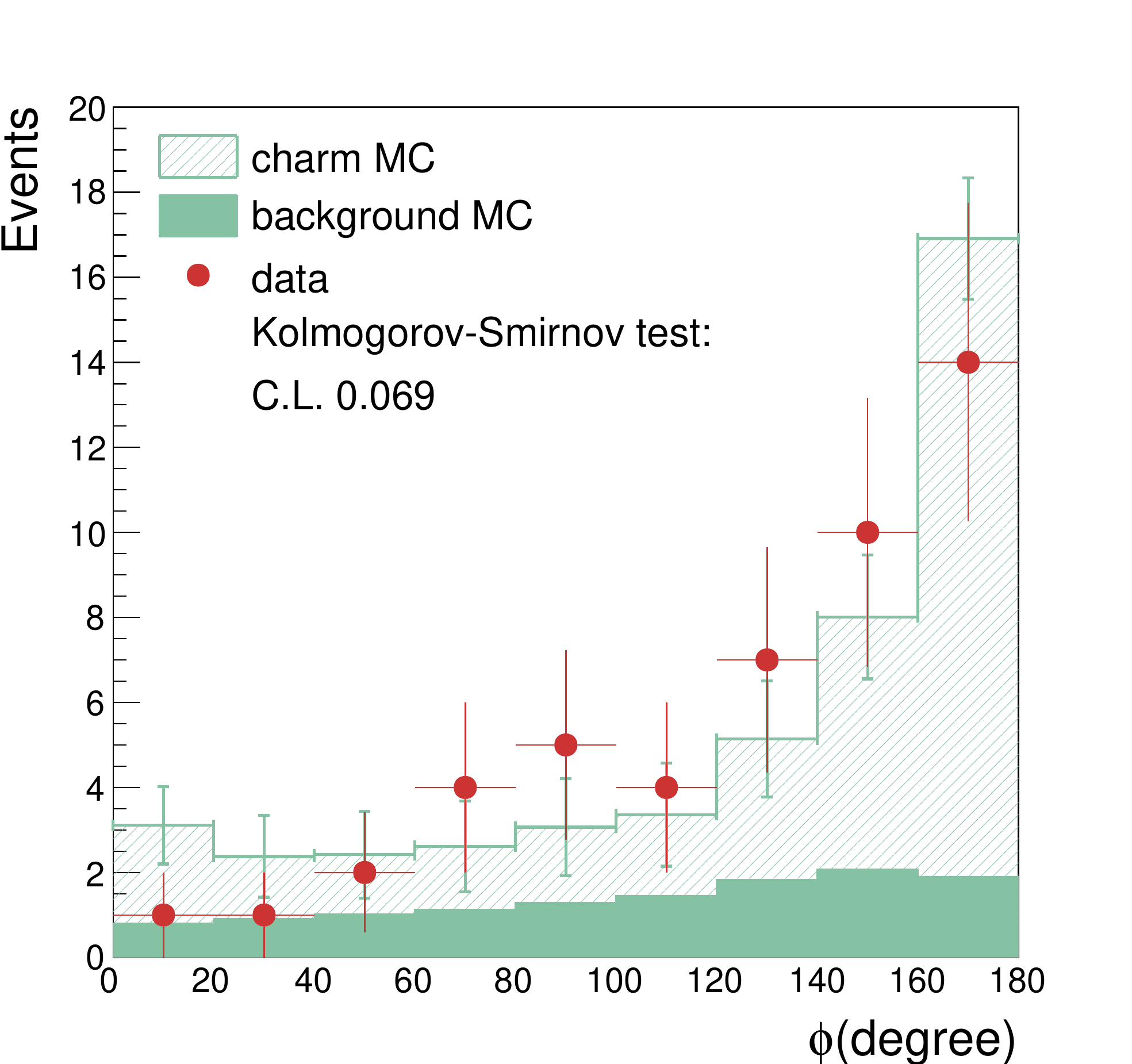}
\end{tabular}
\caption{Shape comparison between observed $\nu_\mu$ CC interactions with candidate charm decays and MC expectations. Left: distribution of the decay length of the candidate charmed particles. Right: distribution of the angle between the candidate charmed particle and the primary muon in the $\nu$ transverse plane. The expected background contribution is also shown (stacked histogram).} \label{fig:charm_FL_PHI}
\end{figure*}

\begin{figure*}[h!]
\centering
\begin{tabular}{cc}
	\includegraphics[scale=0.32]{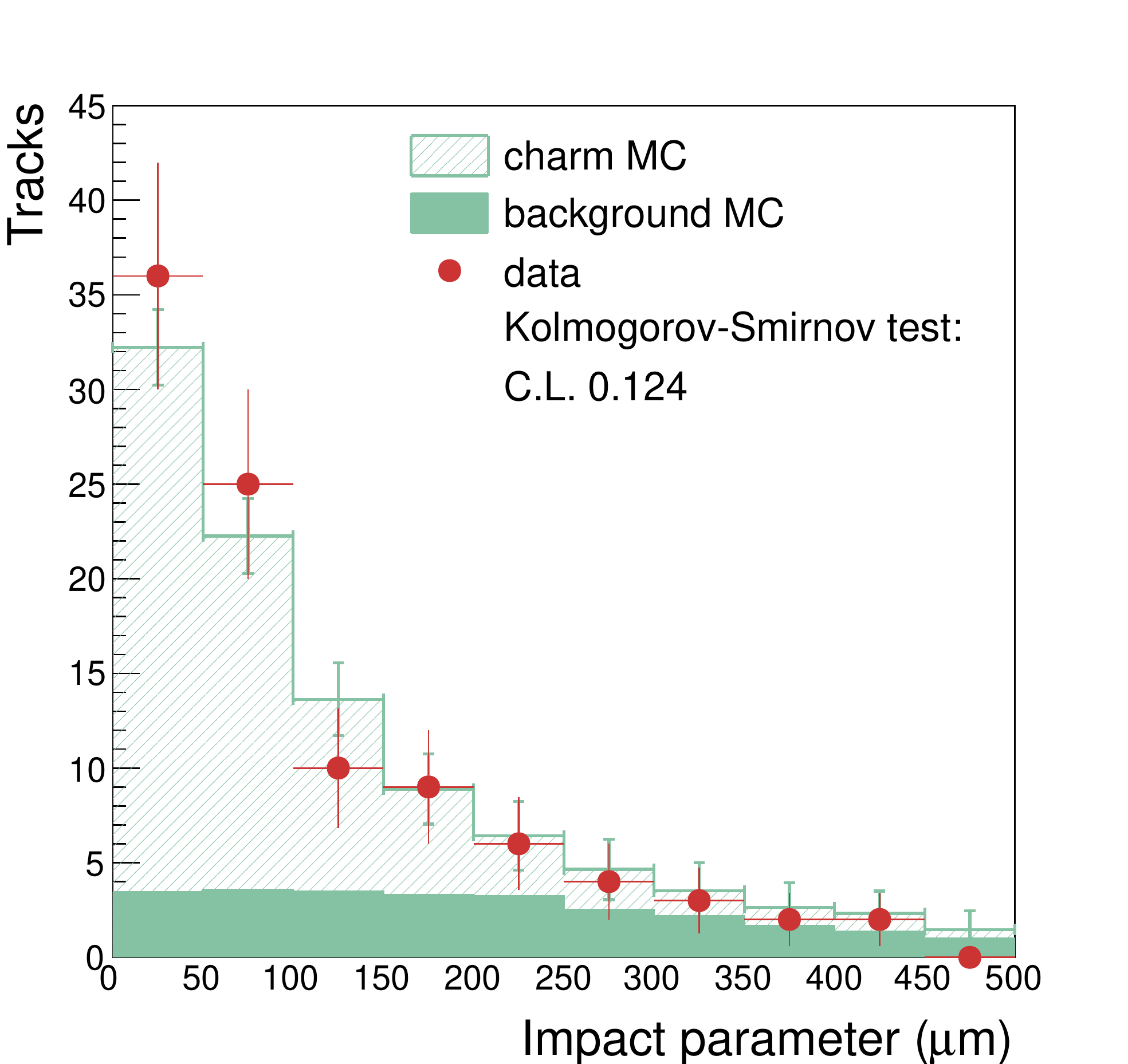}
 & \includegraphics[scale=0.32]{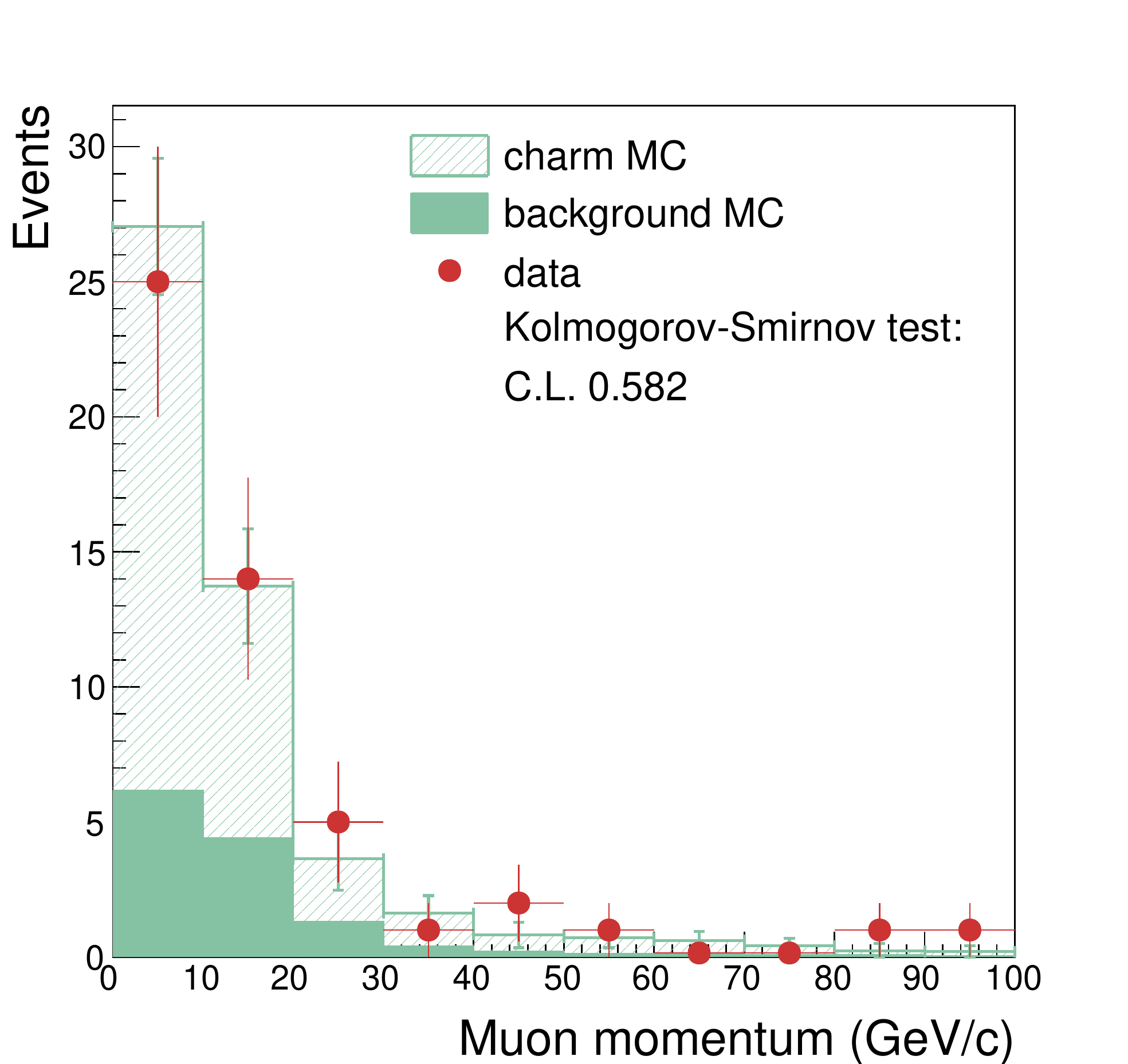}
\end{tabular}
\caption{Shape comparison between observed $\nu_\mu$ CC interactions with candidate charm decays and MC expectations. Left: distribution of the impact parameters of the candidate charm  daughter particles with respect to the neutrino interaction vertex. Right: distribution of muon momentum. The expected background contribution is also shown (stacked  histogram).} \label{fig:charm_IP_PMU}
\end{figure*}

Observed candidate charm events are compared to MC expectations for signal and background in Figures \ref{fig:charm_FL_PHI} and \ref{fig:charm_IP_PMU}.

Figure \ref{fig:charm_FL_PHI} left shows the distribution of the flight length of the candidate charmed particles for data,  the expected charm yield and the estimated background.
The distribution of the angle between the candidate charmed particle and the primary muon in the $\nu$ transverse plane is presented in Figure \ref{fig:charm_FL_PHI} right.

Figure \ref{fig:charm_IP_PMU} left shows the impact parameters of the candidate charm daughter particles with respect to the neutrino interaction vertex.
The distribution of the muon momentum, as measured by the electronic detectors, is displayed in Figure \ref{fig:charm_IP_PMU} right.

The error bars on the expectations shown in Figures \ref{fig:charm_FL_PHI} and \ref{fig:charm_IP_PMU} represent systematic and statistical uncertainties added in quadrature.

The results of the Kolmogorov-Smirnov test applied to the above plots and shown in the legenda indicate that data and MC agree at a confidence level $\gtrsim  \, 10 \, \%$.

The shape comparison illustrated by the plots, as well as the absolute yield comparison reported in Table \ref{tab:charm_res} show that the description of the detector,  of its performance and of the procedures applied in the analysis are correctly reproduced by the simulation. 
Despite the small statistics, the results obtained for the 
charm control sample are crucial to understand the detection efficiency for the $\tau$ lepton showing similar decay topologies.

\begin{figure*}[h!]
\begin{center}
    \includegraphics[scale=0.35]{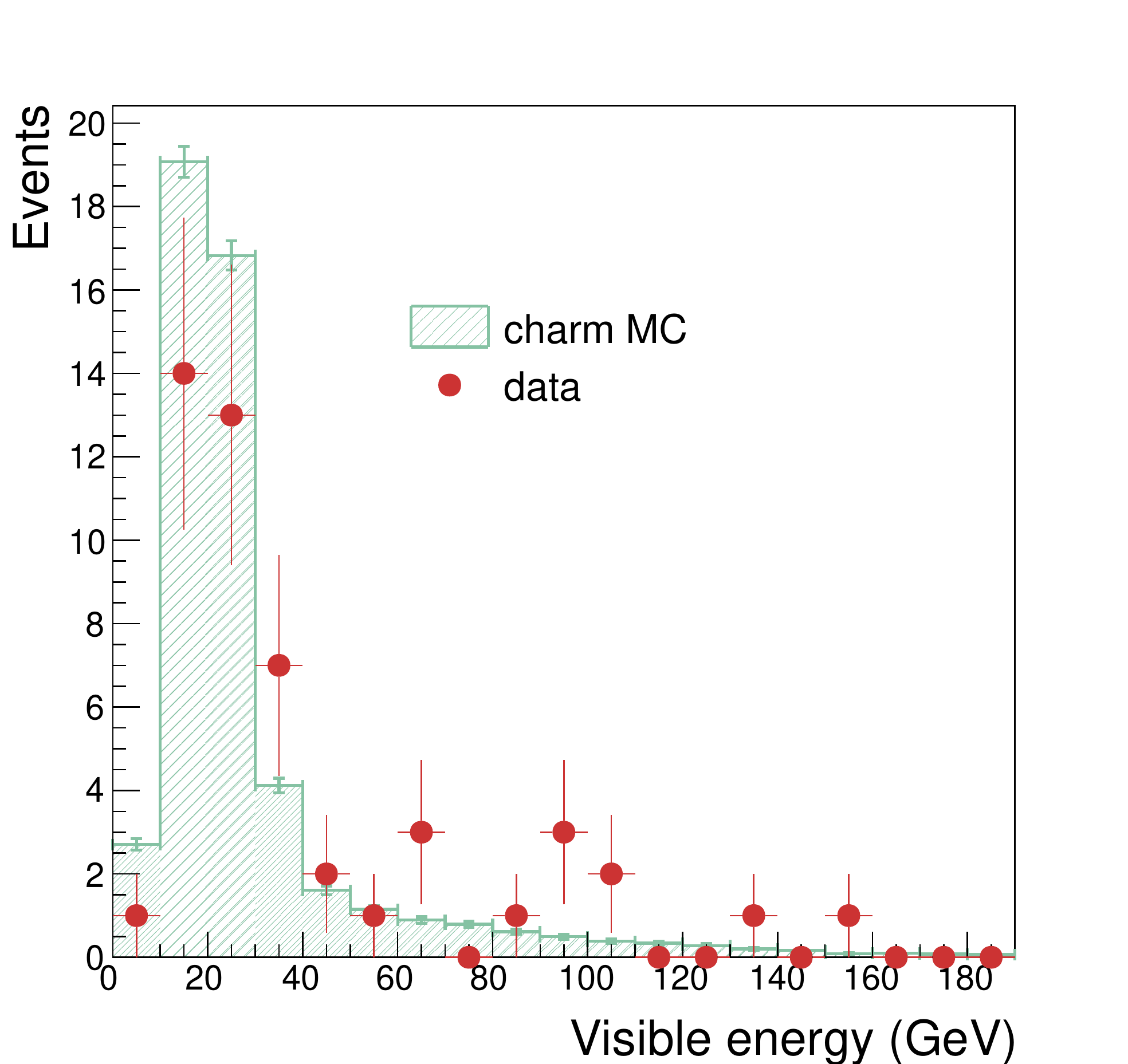}
\end{center}
\caption{Visible energy distribution for observed $\nu_\mu$ CC interactions with candidate charm decays and MC expectations. \label{fig:charm_Evis}}
\end{figure*}

As a further comparison, the distribution of the visible energy, computed as the sum of the hadronic energy deposited in the TT  and the energy of the muon, is presented in Figure \ref{fig:charm_Evis}: data fairly reproduce the expected MC behaviour.

\section{Conclusions}

In the OPERA experiment, the decay search analysis is systematically applied to all located neutrino events to search for oscillated $\nu_\tau$ CC interactions through the direct observation of the $\tau$  lepton decays. The procedure also allows detecting $\nu_\mu$ CC interactions with charmed hadrons in the final state. 
Charmed particles constitute indeed a precious control sample to cross-check the topological $\tau$ detection efficiency, 
due to the very similar decay patterns.  

In this paper, the results obtained in the analysis of the full data sample of located $\nu_\mu$ CC events collected in 2008 and 2009 runs as well as a subsample of the statistics of the 2010 run are reported.

The charm yield is evaluated using CHORUS data after re-weighting for the CNGS neutrino spectrum. 

In total, ($\rm 40 \, \pm 3$) charm and ($\rm 14 \, \pm 3$)  background events, due to hadronic re-interactions and strange particle decays, are expected in the analysed sample, while 
$50$ decay candidate events are observed. 

The comparison between observed and simulated data for several relevant variables proves that the detector performance and the full analysis chain applied to neutrino events are well reproduced by the OPERA simulation chain. 

\begin{acknowledgements}
We thank CERN for the successful operation of the CNGS facility and INFN for the continuous support given to the experiment through its LNGS laboratory. We acknowledge funding from our national agencies. Fonds de la Recherche Scientique-FNRS and Institut InterUniversitaire des Sciences Nucl\éaires
for Belgium, MoSES for Croatia, CNRS and IN2P3 for France, BMBF for Germany, INFN for Italy, JSPS 
(Japan Society for the Promotion of Science), MEXT 
(Ministry of Education, Culture, Sports, Science and 
Technology), QFPU (Global COE programme of Nagoya University, Quest for Fundamental Principles in the Universe supported by JSPS and MEXT) and Promotion and Mutual Aid Corporation for Private Schools of Japan for Japan, SNF, the University of Bern for Switzerland, the Russian Foundation for Basic Research (grant no. 19-02-00213-a, 12-02-12142 ofim), the Programs of the Presidium of the Russian Academy 
of Sciences Neutrino physics and Experimental and theoretical researches of fundamental interactions connected 
with work on the accelerator of CERN, the Programs 
of Support of Leading Schools (grant no. 3110.2014.2), 
and the Ministry of Education and Science of the Russian 
Federation for Russia and the National Research Foundation 
of Korea Grant No. 2011-0029457.
We are also indebted to INFN for providing fellowships and grants to non-Italian researchers. We thank the IN2P3 Computing Centre (CC-IN2P3) for providing computing resources for the analysis and hosting the central database for the OPERA experiment.
We are indebted to our technical collaborators for the excellent quality of their work over many years of design, prototyping, construction and running of the detector and of its facilities.
\end{acknowledgements}

\clearpage

\clearpage

\end{document}